\documentclass[a4paper,aps, prb,amsmath,amssymb, superscriptaddress, showpacs, twocolumn, floatfix]{revtex4-1}
\usepackage{graphicx, braket, mathrsfs, nicefrac, epsfig, multirow, ulem, verbatim, subfigure}

\newcommand{\Mab}[4]{\left(\begin{array}{cc}
#1\ & #2 \\
#3\ & #4 \end{array} \right)}

\begin{document}
\title{Closing the proximity gap in a metallic Josephson junction between three superconductors}
\author{C. Padurariu}
\affiliation{Univ. Grenoble Alpes, Inst. NEEL, F-38000 Grenoble, France.}
\affiliation{CNRS, Inst. NEEL, F-38000 Grenoble, France.}
\author{T. Jonckheere} 
\affiliation{Aix-Marseille Universit\'e, Universit\'e de Toulon, CNRS, CPT, UMR 7332, 13288 Marseille, France.}
\author{J. Rech}
\affiliation{Aix-Marseille Universit\'e, Universit\'e de Toulon, CNRS, CPT, UMR 7332, 13288 Marseille, France.}
\author{R. M\'elin}
\affiliation{Univ. Grenoble Alpes, Inst. NEEL, F-38000 Grenoble, France.}
\affiliation{CNRS, Inst. NEEL, F-38000 Grenoble, France.}
\author{D. Feinberg} 
\affiliation{Univ. Grenoble Alpes, Inst. NEEL, F-38000 Grenoble, France.}
\affiliation{CNRS, Inst. NEEL, F-38000 Grenoble, France.}
\author{T. Martin} 
\affiliation{Aix-Marseille Universit\'e, Universit\'e de Toulon, CNRS, CPT, UMR 7332, 13288 Marseille, France.}
\author{Yu. V. Nazarov}
\affiliation{Kavli Institute of Nanoscience, Delft University of Technology, Lorentzweg 1, 2628 CJ, Delft, The Netherlands.}

 \begin{abstract}
 
We describe the proximity effect in a short disordered metallic junction between three superconducting leads. 
Andreev bound states in the multi-terminal junction may cross the Fermi level. We reveal that for a quasi-continuous metallic density of states, crossings at the Fermi level manifest as closing of the proximity-induced gap. 
We calculate the local density of states for a wide range of transport parameters using quantum circuit theory.
The gap closes inside an area of the space spanned by the superconducting phase differences.
We derive an approximate analytic expression for the boundary of the area and compare it to the full numerical solution. 
The size of the area increases with the transparency of the junction and is sensitive to asymmetry.
The finite density of states at zero energy is unaffected by electron-hole decoherence present in the junction, 
although decoherence is important at higher energies. Our predictions can be tested using 
tunneling transport spectroscopy. To encourage experiments, we calculate the current-voltage 
characteristic in a typical measurement setup. We show how the structure of 
the local density of states can be mapped out from the measurement.

 \end{abstract}

\pacs{ Pacs }
\maketitle

\section{Introduction}
\label{sec:introduction}



The density of states in a normal metal is strongly modified by contact to one or multiple superconductors placed in its proximity. 
In junctions with two superconductors, the development of a minigap in the density of states has been in the focus of research for 
many years \cite{Golubov}. Recent attention has been given to the possibility of engineering states inside the minigap in the context 
of Majorana zero energy states \cite{Kitaev, Fu}. To this end, it would be advantageous if the Andreev states in a Josephson junction could 
be brought to zero energy, i.e. at the Fermi level, by controlling only the superconducting phase of the junction.
Zero energy Andreev states together with spin-orbit coupling provide the opportunity to manipulate single fermionic quasiparticles \cite{Anton}. The motivation is to
increase understanding of quasiparticle dynamics and facilitate designs of superconducting spin quantum memory bits \cite{meandYuli}. 
Unfortunately, in two terminal Josephson junctions Andreev states do not cross the Fermi level except at a singular phase difference \cite{Carlo, Schon},  $\varphi=\pi$, 
and only in systems with reflectionless transport channels.

\begin{figure}
\centerline{\includegraphics[width=0.9\linewidth]{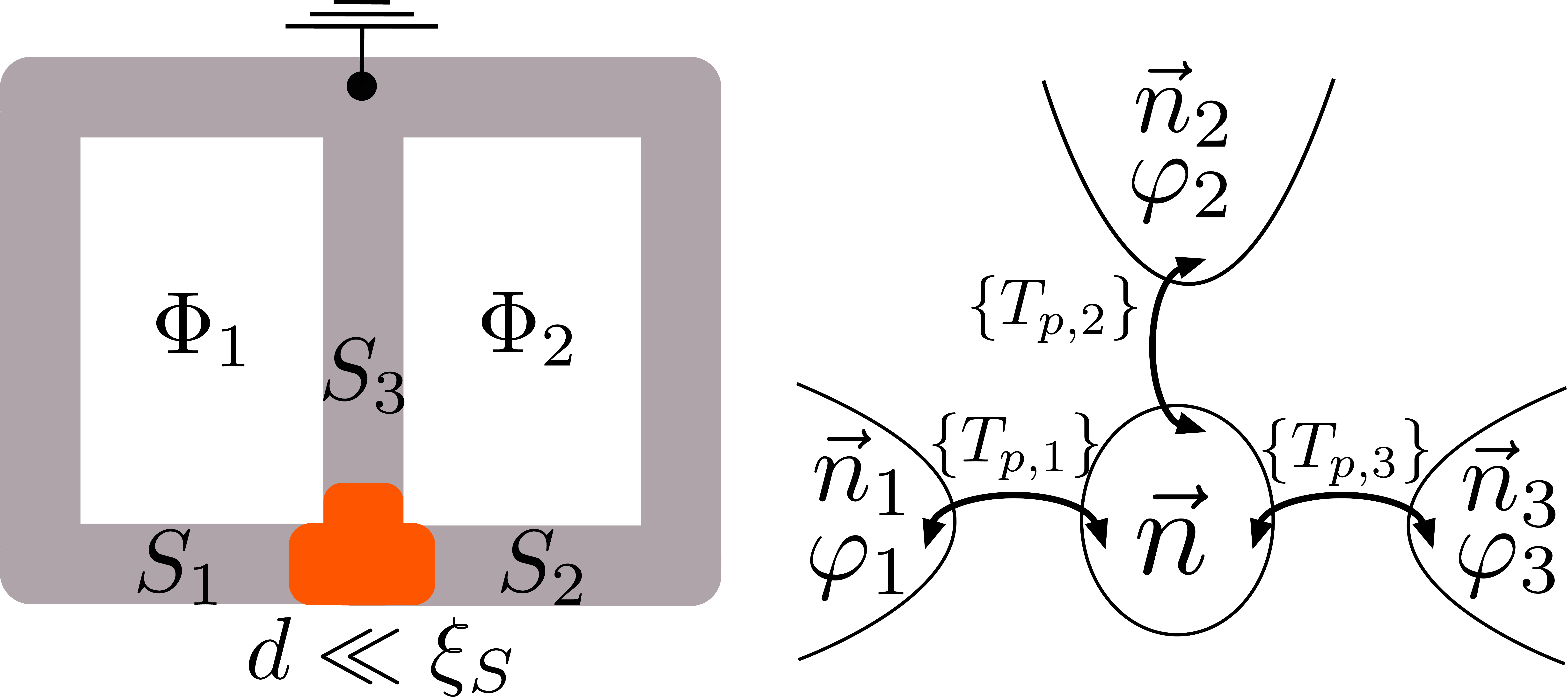}}
\caption{(Color online.) \textit{Left.} Two superconducting rings can be used to control the two independent superconducting 
phase differences in a three terminal superconducting junction via fluxes $\Phi_{1,2}$. 
\textit{Right.} Quantum circuit theory model of a three terminal superconduting device using a single node. 
}
\label{fig:model}
\end{figure}



Recent work \cite{Anton, Riwar, Tomohiro} shows that the situation is different in a Josephson junction with multiple terminals ($N \geq 3$). 
The Andreev states have been shown to cross the Fermi level if the superconducting phases wind by $2\pi$ around the junction \cite{Anton}. The crossing points have been termed Weyl points, as they are analogous to Weyl singularities studied in 3D solids, with Andreev bound states corresponding to energy bands and the superconducting phase differences corresponding to quasi-momenta \cite{Riwar, Tomohiro}.
The presence of Weyl points has been illustrated 
theoretically by using scattering theory to model junctions consisting of a 
quantum dot connected to three \cite{Anton} or four \cite{Riwar, Tomohiro} superconductors. In particular, in three-terminal junctions, Weyl points appear along a closed curve in the space spanned 
by two independent superconducting phases. 
The area enclosed by the curve is shown to be maximal for a fully transparent quantum dot \cite{Anton}.

\begin{figure}
\centerline{\includegraphics[width=0.9\linewidth]{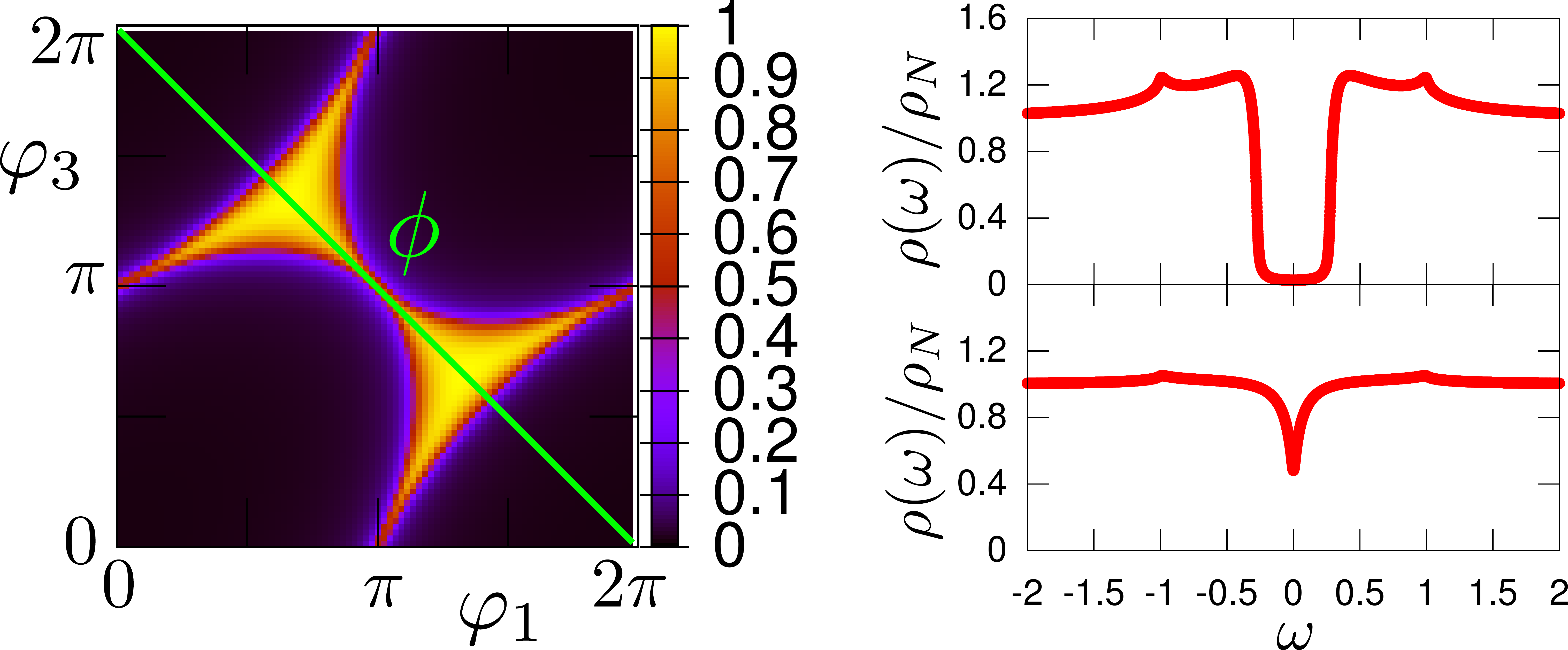}}
\caption{(Color online.) 
\textit{Left.} Local density of states at the Fermi level as a function of phase differences for 
a symmetric junction with transmission coefficient $T=0.9$. 
The figure illustrates the gapped regime (dark colored regions) and gapless regime (light colored regions). 
We define the phase $\phi=\varphi_1=-\varphi_3$, with $\varphi_2=0$, that parameterizes an axis along which
the system undergoes transitions between gapped and gapless regimes.
\textit{Right.} Above: the density of states as a function of energy in the gapped regime, at $\phi=0.44\pi$. 
Below: the same in the gapless regime, at $\phi=0.63\pi$.
}
\label{fig:dos}
\end{figure}


Weyl points have been found for some choices of the scattering matrix describing the junction, but not for all. The existence or absence of Weyl points in the spectrum is determined by the details of the junction scattering matrix. 
The requirements for presence of Weyl points are not yet well understood, but are very important in view of their experimental 
observation. From the viewpoint of the experimental realization, it is advantageous to study metallic junctions, 
that do not require the strict electrical confinement of quantum dots. 
Metallic junctions have a quasi-continuous spectrum of states, in contrast to the discrete states in a quantum dot. 
The larger number of states discourages use of the scattering theory approach. To model metallic junctions, quasiclassical methods are better suited.

\begin{figure*}
\centerline{\includegraphics[width=0.7\linewidth]{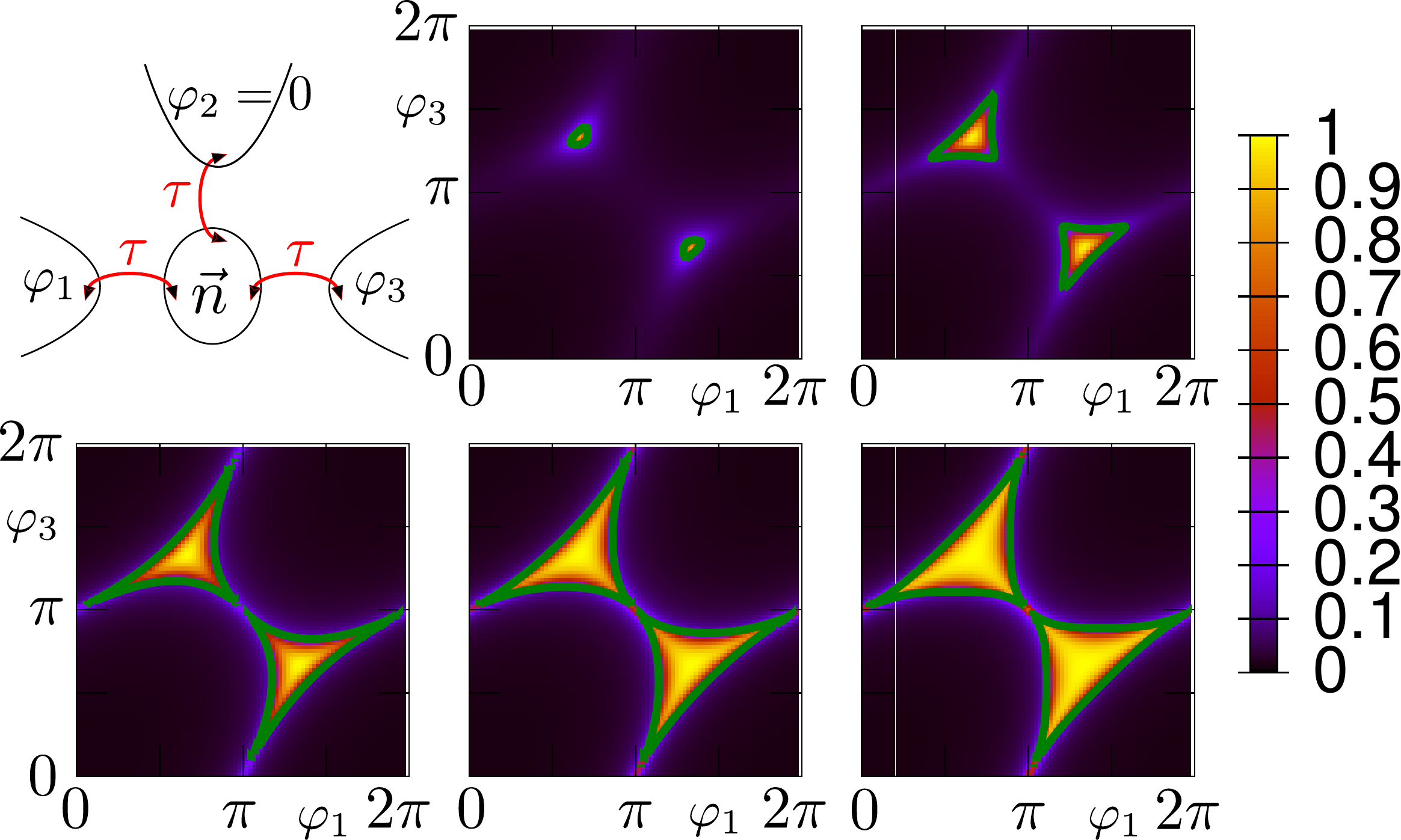}}
\caption{(Color online.) \textit{Top left:} Schematic of a fully symmetric junction with transmission coefficient $\tau$.
\textit{Left to right:} The local density of states is calculated at the Fermi level as a function of phases,
for transmissions $\tau=0.1, 0.3, 0.5, 0.7, 0.9$, using the full numerical solution. The horizontal $x$-axis represents $\varphi_1$,
while the vertical $y$-axis represents $\varphi_3$, with $\varphi_2=0$.
The boundary of the gapless region calculated using the result in Eq. \ref{eq:solution} is shown in bright green.
}
\label{fig:dosatzerosymm}
\end{figure*}
In this paper, we study the local density of states in a disordered metallic junction between three superconductors (see Fig. \ref{fig:model}). 
The dimensions of the junction are assumed short at the scale of the superconducting coherence length, but much larger 
than the mean free path of electrons in the metallic island, in the normal state. We show that the junction exhibits 
two regimes. One where the density of states has a finite minigap, similar to the minigap in two terminal junctions, 
and another where the minigap is closed and the density of states is finite at all energies. The transition between these 
regimes can be made in equilibrium, by varying the superconducting phase differences. The absence of the minigap 
is a manifestation of states crossing at the Fermi level. In metallic junctions the gapless regime 
corresponds to an area in the space spanned by 
the superconducting phase differences, as in Fig. \ref{fig:dos}, in 
contrast to quantum dot junctions where the regime corresponds only to a closed curve. The larger parameter 
space is an additional advantage of metallic junctions in view of the experimental observation of the effect.

The paper is organized as follows. In Sec. \ref{sec:theoretical} we introduce the theoretical method used. Sec. \ref{sec:analytical} presents our analytical derivation of the local density of states in the junction, showing the presence of a gapped and a gapless regime. Sec. \ref{sec:numerical} presents the comparison between analytical and numerical results, discussing the physical mechanism at the origin of the gapless regime. Sec. \ref{sec:decoherence} discusses the effect of electron-hole decoherence in the junction. In Sec. \ref{sec:measurement} we propose to observe the predictions using tunneling spectroscopy and provide a numerical simulation of the results of a typical measurement setup.
Sec. \ref{sec:conclusions} presents our conclusions.



\section{Theoretical method}
\label{sec:theoretical}

To determine the local density of states in the disordered metallic island, we use the method of quantum circuit theory \cite{Yuli}, 
whereby the junction is separated into circuit elements described by spatially independent quasiclassical Green's functions. 
If the size of the island is much smaller than the superconducting coherence length,  the appropriate circuit contains a single node 
connected to three superconducting terminals (see Fig. \ref{fig:model}).
Transport in the circuit is described using the quasiclassical action that is extremized with respect to the unknown retarded Green's 
function in the island $\hat{G}$.
The retarded Green's function matrix is conveniently parametrized by $\hat{G} = (\vec{n}(\theta,\phi)\cdot\hat{\vec{\tau}})$, 
where $\hat{\vec{\tau}}$ is the vector of Pauli matrices in Nambu space. The spectral vector \cite{Yulivectors}, $\vec{n}(\theta,\phi)$, 
is generally parametrized by two complex numbers $\theta$ and $\phi$, with  
$n_x = \sin(i\theta)\sin(\phi)$, $n_y = \sin(i\theta)\cos(\phi)$, and $n_z = \cos(i\theta)$. 
\begin{figure*}
\centerline{\includegraphics[width=0.7\linewidth]{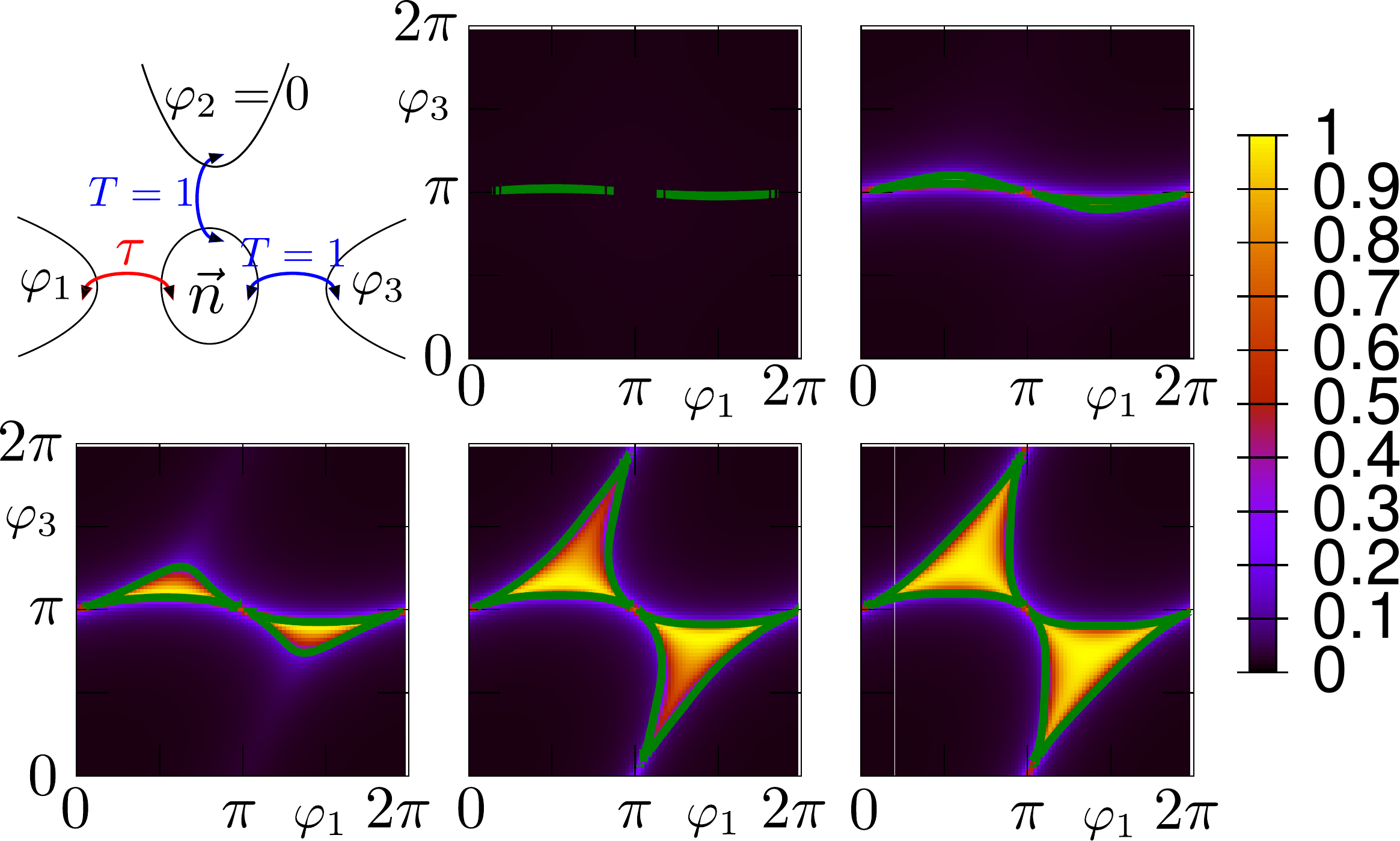}}
\caption{(Color online.) \textit{Top left:} Schematic of an asymmetric junction with one contact described by transmission $\tau$ and the other two fully transparent.
\textit{Left to right:} The same as in Fig. \ref{fig:dosatzerosymm} for the asymmetric junction with two fully transparent contacts.
}
\label{fig:dosatzeroasymm1}
\end{figure*}
In the superconducting bulk with energy gap $|\Delta|$, assumed the same for all three superconductors, 
$\phi_S$ takes the value of the superconducting phase 
$\varphi_i$ corresponding to superconductor $i=\{1,2,3\}$, and $\theta_S$ has the following energy dependence,
\begin{align}
\theta_S = \begin{cases} -i\frac{\pi}{2}+\frac{1}{2}\ln\frac{1+\omega}{1-\omega} , & \omega<1 , \\
                                         \frac{1}{2}\ln\frac{\omega+1}{\omega-1} , &  \omega>1 ,   \end{cases}
\notag
\end{align}
where $\omega = E/|\Delta|$. In the limit of low energy $\omega \ll 1$, the spectral vector in superconductor $i$ has the following simple form,
$\vec{n}_i = (\sin(\varphi_i),\cos(\varphi_i),-i\omega)$.
In terms of the spectral vector of the island $\vec{n}$ and of those of the superconductors $\vec{n}_i$, the quasiclassical action of the junction takes the form
\begin{align}
S = 2{\rm Re}\left\{\sum_i\sum_p \ln\left[1+\frac{T_{p,i}}{2}\left((\vec{n}\cdot\vec{n}_i)-1\right)\right] \right\},
\label{eq:action}
\end{align}
where $p$ labels the open transport channels in each of the junction contacts and $T_{p,i}$ represents the corresponding transmission coefficient. 
The spectral vector in the metallic island $\vec{n}(\theta,\phi)$ is obtained by extremization of the action under the constraint $\vec{n}^2=1$. 
We introduce a complex Lagrange multiplier $\lambda$ such that the extremization problem reduces to a set of four complex equations
\begin{align}
\vec{\nabla}_{\vec{n},\lambda} \Lambda(\vec{n}, \lambda) = \vec{0}, \quad \Lambda(\vec{n},\lambda)=S(\vec{n})+\lambda \vec{n}^2.
\label{eq:balance}
\end{align}
In general, the equations are strongly non-linear in terms of the unknowns, $\vec{n}$ and $\lambda$, and are solved numerically. 
The local density of states in the metallic island is obtained from $\vec{n}$ by 
$\rho(\omega)/\rho_N={\rm Re}\{n_z\}={\rm Re}\{\cos(i\theta)\}$, where $\rho_{N}$ is the density of states in the normal state.

\begin{figure*}
\centerline{\includegraphics[width=0.7\linewidth]{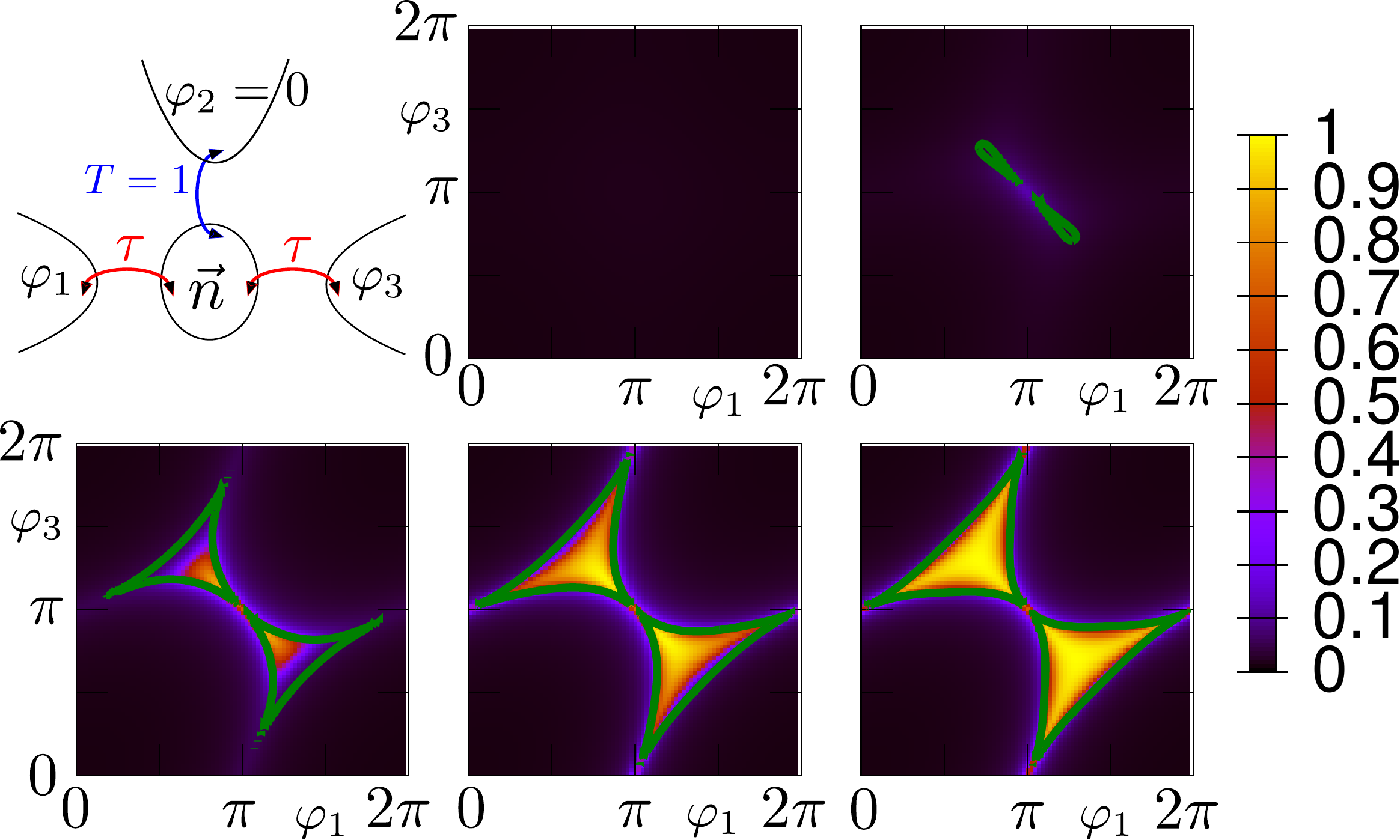}}
\caption{(Color online.) \textit{Top left:} Schematic of an asymmetric junction with two contacts described by transmission $\tau$ and the third fully transparent.
\textit{Left to right:} The same as in Fig. \ref{fig:dosatzerosymm} for the asymmetric junction with a single fully transparent contact.
}
\label{fig:dosatzeroasymm2}
\end{figure*}
\section{Analytical Results}
\label{sec:analytical}

Before we discuss the full numerical solution, we present analytic results in the tunnel limit, i.e. the transmission 
coefficients of all channels are small, $T_{p,i}\ll 1$. In the lowest order, the action takes the form 
\begin{align}
S^{(1)} = 2{\rm Re}\{(\vec{n}\cdot \vec{n}_s)\}, 
\label{eq:action1}
\end{align}
where $\vec{n}_s=\sum_i g_i\vec{n}_i$ is a spectral vector that depends on the superconducting phase differences 
and $g_i=\sum_p (T_{p,i}/2)$ is proportional to the normal state conductance of contact $i$ in units of the 
conductance quantum. We find two types of solutions in terms of $\vec{n}_s$ that differ in the structure of the 
local density of states. i. If $\vec{n}_s$ is non-vanishing, the extrema of the action $S^{(1)}$ occur when the 
vectors $\vec{n}$ and $\vec{n}_s$ are aligned. In this case, we find that the z-component of $\vec{n}$ is purely 
imaginary in the limit of low energy $\omega\ll 1$, $n_z = -i\omega$, meaning that the local density of states 
in the island has a finite gap. ii. In the opposite case, when $\vec{n}_s=\vec{0}$, the action vanishes in the 
lowest order. To study this situation we include the next order terms in the action 
\begin{align}
S^{(2)} = 2{\rm Re}\left\{\sum_i \left[(g_i+q_i)(\vec{n}\cdot\vec{n}_i) - \frac{q_i}{2} (\vec{n}\cdot\vec{n}_i)^2\right]\right\},
\label{eq:action2}
\end{align}
where $q_i=\sum_p \left(T_{p,i}/2\right)^2$ is proportional to the conductance of coherent processes that transfer 
two pairs of quasiparticles between the leads. 
We solve Eq. \ref{eq:balance} using the action $S^{(2)}$, in the limit of low energy $\omega \ll 1$,
\begin{align}
{\rm Re}\left\{\sum_i \left[(g_i+q_i) - q_i (\vec{n}\cdot\vec{n}_i)\right]\vec{n}_i\right\}+\lambda\vec{n}=0,
\label{eq:balance2}
\end{align}
together with $n_z = \sqrt{1-n_x^2-n_y^2}$. The $z$-component of the equation is of order $\omega$ and can be 
eliminated if the Lagrange multiplier $\lambda$ is also of order $\omega$. The $x$- and $y$-components of the 
equation are of order $1$ and can be solved for $n_{x,y}$,
\begin{align}
\ &n_x = \frac{a_4a_3-a_1a_5}{2\left(a_3^2-a_2a_5\right)} ; \quad n_y = \frac{a_1a_3-a_4a_2}{2\left(a_3^2-a_2a_5\right)} ;\notag\\
\ &n_z = \sqrt{1-n_x^2-n_y^2} .
\label{eq:solution}
\end{align}
where we define the coefficients $a_1 = \sum_i (g_i+q_i)\sin(\varphi_i)$, $a_2 = \sum_i q_i \sin^2(\varphi_i)$, 
$a_3 = \sum_i q_i \sin(\varphi_i)\cos(\varphi_i)$, $a_4 = \sum_i (g_i+q_i)\cos(\varphi_i)$, and $a_5 = \sum_i q_i \cos^2(\varphi_i)$. 
Components $n_x$ and $n_y$ are real, while component $n_z$ is real if $n_x^2+n_y^2 \leq 1$ and purely imaginary otherwise. 
The relation $n_x^2+n_y^2 \leq 1$ defines an area in the space of superconducting phases where the local 
density of states in the island is finite at small energies, i.e. the gap in the density of states is closed. 
Estimations of $n_x$ and $n_y$ yield $n_{x,y}\propto T^{-1}$, 
meaning that $n_x^2+n_y^2 \leq 1$ defines a smaller area for smaller transmission coefficients. 
We have compared the analytic results for the boundary of this area with our full numerical solution in 
Figs. \ref{fig:dosatzerosymm}, \ref{fig:dosatzeroasymm1}, and \ref{fig:dosatzeroasymm2} and have found excellent 
agreement up to large transmission coefficients.

\begin{figure*}
\centerline{\includegraphics[width=0.7\linewidth]{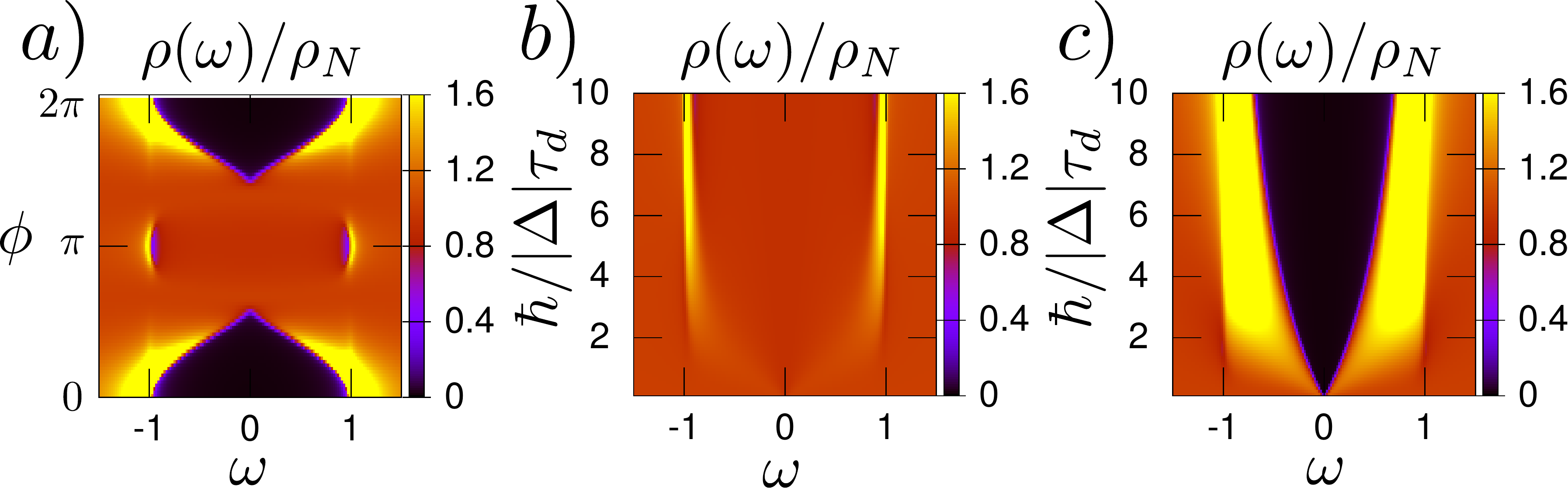}}
\caption{(Color online.) \textbf{a.} Density of states as a function of energy plotted for different values of the phase $\phi$ defined in Fig. \ref{fig:dos}. 
\textbf{b.} Density of states plotted as a function of energy for different dwell times, $\tau_d$, and for $\phi$ chosen in the gapless regime, at $\phi=\pi$.
\textbf{c.} The same as for panel \textbf{b}, for $\phi$ chosen in the gapped regime, at $\phi=0.1\pi$.
\label{fig:decoherence}
}
\end{figure*}

\section{Numerical results}
\label{sec:numerical}
 
The numerical method used to obtain the exact solution is presented in Appendix.
To restrict the number of parameters, we assume in all calculations that the number of open channels 
is the same for all contacts and that the channels in each contact have equal transmission coefficients, 
i.e. they are independent of channel label $p$. We stress that our results are not restricted to this situation; 
the method can be applied for any type of contacts. We have checked that the conclusions we present are
robust with respect to change of the details of the contacts, as well as to change of the relative size of the superconducting
gap of the three superconducting leads.

The details of the contacts determine the shape of the area corresponding to the gapless regime. 
For a symmetric junction this area diminishes, but does not vanish, at low transparency
(see Fig. \ref{fig:dosatzerosymm}). As the transparency is increased the area reaches a maximum at $T=1$. 
It is notable that the maximal area resembles the full parameter space where the superconducting 
phases wind by $2\pi$ around the junction, i.e. the necessary condition for 
presence of Weyl points \cite{Anton}, but does not fully cover it.

For the asymmetric junction with two fully transparent contacts, described in Fig. \ref{fig:dosatzeroasymm1}, 
the case of low transparency corresponds to a symmetric junction with only two, rather than three, superconductors.
The symmetric two terminal junction exhibits reflectionless transport channels leading to a crossing 
at the Fermi level at the singular point $\varphi_3=\pi$. As the transmission of the third contact increases, the line $\varphi_3=\pi$ is deformed gradually and the area corresponding to the gapless regime increases.

In the case of the asymmetric junction with a single fully transparent contact, described in Fig. 
\ref{fig:dosatzeroasymm2}, we find that the presence of a gapless regime is conditioned by 
the transmission coefficient exceeding a threshold value. 
In contrast, for the symmetric junction we find a non-vanishing area corresponding to the gapless regime at all low transparencies. 
This denotes an apparent significance of processes involving all three superconductors 
in the emergence of the gapless regime.


Let us discuss the physical mechanism that leads to closing of the minigap in the density of states. 
The gapless regime corresponds to the situation when
the balance of currents is dominated by transport processes involving 
coherent transfer of multiple Cooper pairs between the leads. 
This follows from our analytic derivation where second order terms
are needed to find extrema of the action in perturbation theory. 
Multi-pair processes are not special to the geometry with three superconductors, they also
manifest in conventional Josephson junctions with two superconductors. However,
in conventional junctions multi-pair processes do not
give rise to a gapless regime \cite{Gueron}. 
It is possible that the gapless regime emerges as a result of non-local multi-pair processes that 
require multiple terminals ($N \geq 3$). Non-local transport processes consist of exchanges 
of Cooper pairs that involve quasiparticles from at least three superconductors \cite{Freyn, Jerome, Denis}. 
The superconducting current contributed by non-local transport processes depends simultaneously on all 
three superconducting phases involved. 

The fundamental process at the origin of non-local transport is the non-local Andreev reflection, 
also termed crossed Andreev reflection. Crossed Andreev reflection has been studied for years 
in the context of junctions formed between a superconductor and two normal leads \cite{Becks, Falci, Belzig}. 
On the contrary, the physics of crossed Andreev reflection in junctions with three superconducting leads 
has received relatively little attention until now. Recent extensive studies have so far been restricted 
to transport through a single level dot \cite{Jerome, Denis}. 
The rigorous analysis of non-local processes 
in metallic junctions with three superconductors is beyond the scope of this 
paper and will be reported elsewhere.

\section{Decoherence}
\label{sec:decoherence}

Let us turn our attention to the effect of electron-hole decoherence that appears as a result of quasiparticles spending 
a finite dwell time, $\tau_d$, in the junction. Within quantum circuit theory, decoherence is modeled by adding the following 
term to the action that corresponds to a fictitious circuit element where electron-hole coherence can be dissipated \cite{Yulibook} (see Fig. \ref{fig:model}), 
\begin{align}
S_f = 2{\rm Re}\left\{\sum_i\sum_p \ln\left[1+\frac{T_{p,i}}{2}\left((\vec{n}\cdot\vec{n}_f)-1\right)\right] \right\},
\end{align}
where $\vec{n}_f=(|\Delta|\tau_d/\hbar) (0,0,-i\omega)$.
This additional term modifies only the $z$-component of the vector equation $\vec{\nabla}_{\vec{n}} \Lambda(\vec{n}, \lambda) = \vec{0}$ 
that can be eliminated at low energy $\omega\ll 1,(\hbar/|\Delta|\tau_d)$. It does not modify the $x$- and $y$- components, 
leaving the low energy density of states unchanged. Thus, the boundary of the area in the space of superconducting phases 
where the gap is absent remains unaffected by decoherence. 

On the contrary, away from the limit of low energy, $\omega\gtrsim \min(1,(\hbar/|\Delta|\tau_d))$, 
decoherence plays an important role in the structure of the density of states, as illustrated in Fig. \ref{fig:decoherence}. 
In the gapless regime, the minigap is replaced by a dip in the density of states. As illustrated in Fig. \ref{fig:decoherence}b, the effect of decoherence is to decrease the width of this dip. The dip narrows more rapidly when $\hbar/(|\Delta|\tau_d) \lesssim 1$.
In the gapped regime, we recover a similar effect of decoherence as in the two 
terminal junctions, where the size of the minigap depends strongly on $\hbar/(|\Delta|\tau_d)$ (see Fig. \ref{fig:decoherence}c). 

Near $\phi=\varphi_1=-\varphi_3=\pi$, the density of states exhibits a smile-shaped structure located just below the superconducting gap (see Fig. \ref{fig:decoherence}a), 
reminiscent of the secondary smile-shaped minigap predicted previously in two-terminal junctions \cite{smilegap}. We note that the structure 
is present at low decoherence, but vanishes for intermediate decoherence, when $\hbar/(|\Delta|\tau_d)\simeq 1$ 
(see Fig. \ref{fig:decoherence}b). Further investigation of this structure will be presented elsewhere.


\begin{figure}
\centerline{\includegraphics[width=0.9\linewidth]{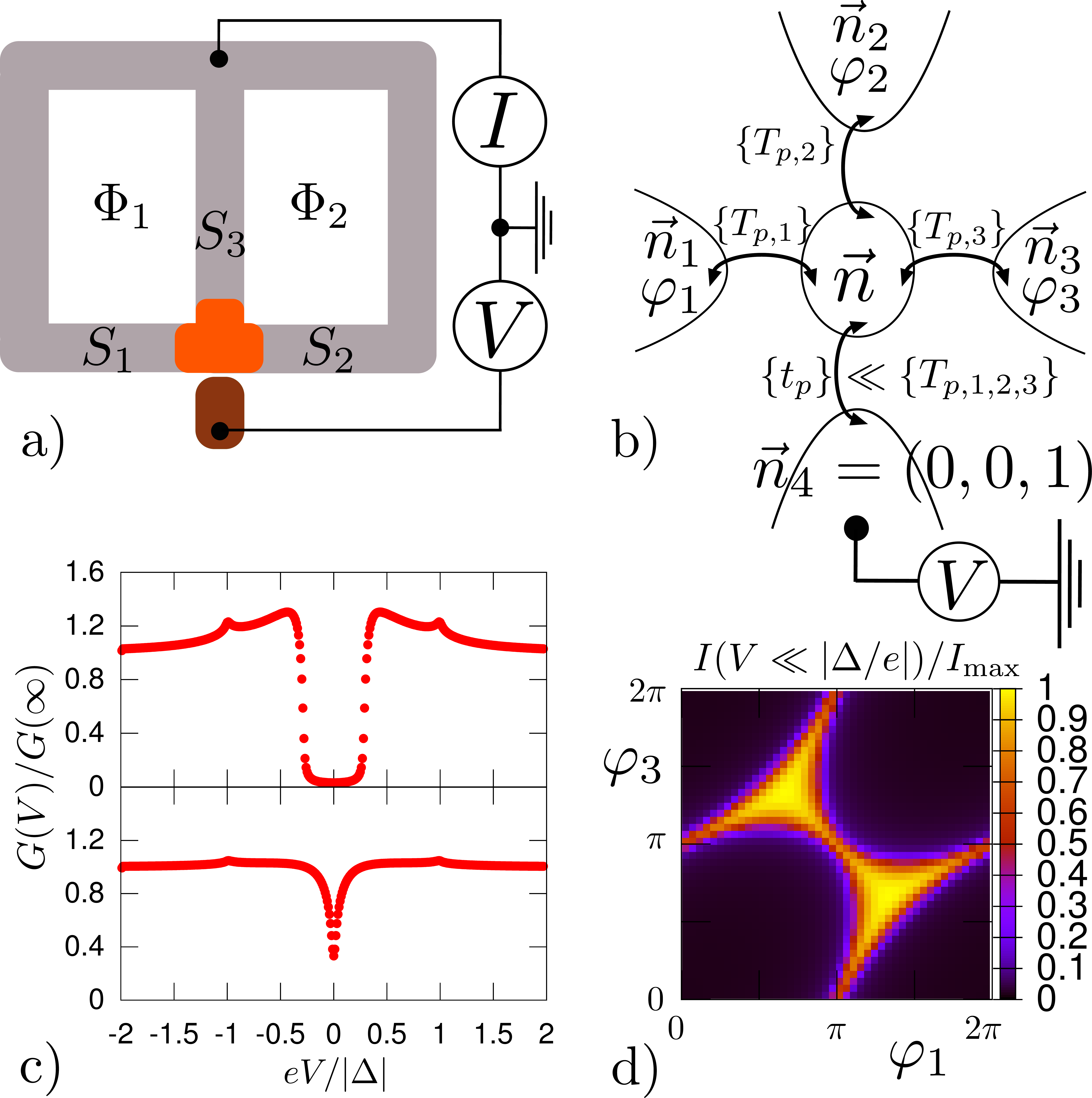}}
\caption{(Color online.) \textbf{a.} Schematic of the tunneling transport spectroscopy measurement setup.
\textbf{b.} Quantum circuit theory model of the measurement setup.
\textbf{c.} Differential conductance of the tunnel probe, $G=dI/dV$, plotted as a function of voltage for the same parameters as in Fig. \ref{fig:dos}.
\textbf{d.} Current measured at low voltage, $eV\ll |\Delta|,\hbar\tau_d^{-1}$, plotted as a function of phases for the same parameters as in Fig. \ref{fig:dos}. The current is normalized by the maximum current at low voltage, $I_{\rm max}$, obtained at $(\varphi_1,\varphi_3)=(2\pi/3,4\pi/3)$ and $(4\pi/3,2\pi/3)$, with $\varphi_2=0$.
\label{fig:measurement}
}
\end{figure}

\section{Measurement}
\label{sec:measurement}

Our predictions can be observed experimentally by means of tunneling spectroscopy \cite{Gueron, DelftMajorana}. 
The minimal setup has an additional normal metal lead brought in tunnel contact to the junction, 
as in Fig. \ref{fig:measurement}a. The differential conductance of the tunnel 
contact is proportional to the local density of states, 
$\rho(eV)$, measured at the applied voltage, provided the contact is sufficiently weak.
In this case, the local density of states is not significantly perturbed by the presence of the tunnel probe.

To check this requirement for our three-terminal superconducting junction, we have used
the quantum circuit model in Fig. \ref{fig:measurement}b to calculate numerically the tunnel current
as a function of bias voltage. The results obtained are presented such that the normalized quantities
 are independent of the contact transmission coefficient, chosen to $t_p=0.01$ in the numerics.
 
In Fig. \ref{fig:measurement}c we show that the differential conductance of the tunnel contact plotted
as a function of bias voltage reproduces the energy dependence of the density of states plotted in
Fig. \ref{fig:dos}, for the same transport parameters.
The signature of the gapless regime can also be observed by measuring the low bias current 
flowing through the tunnel contact as a function of superconducting phases. Fig. \ref{fig:measurement}d
shows that the low bias current matches the dependence depicted in Fig. \ref{fig:dos}, 
mapping out the local density of states at zero energy as a function of superconducting phase differences.

The plots in Fig. \ref{fig:measurement} have been obtained assuming the limit of vanishing temperature.
Our theoretical approach is also valid for finite temperatures, assuming the temperature is sufficiently 
small to permit a well developed proximity effect, $ k_BT \ll \min\{|\Delta|, \hbar\tau_d^{-1}\} $. 
In this temperature window, the local density of states in the junction is independent of temperature.

However, from the viewpoint of the experimental realization, at finite temperature, the features presented 
in Fig. \ref{fig:measurement}c are blurred by thermal broadening of electronic transport in the normal 
contact at the scale $eV \simeq k_BT$. If $\min\{|\Delta|, \hbar\tau_d^{-1}\}\gg k_BT$, the thermal broadening
does not conceal the transition between the gapped and gapless regimes. The measurement simulated in 
Fig. \ref{fig:measurement}c can be used as a signature of this transition.
The measurement simulated in Fig. \ref{fig:measurement}d does not depend on temperature if the tunnel 
current can be measured at a voltage bias chosen such that $eV \ll \min\{|\Delta|, \hbar\tau_d^{-1}\}$ and 
$eV \gg k_BT$.


\section{Conclusion}
\label{sec:conclusions}

In conclusion, we have shown that the minigap in the density of states in a three terminal metallic junction can be closed 
by varying the superconducting phases. The area in the space of phase differences where the minigap is closed increases with the junction 
transparency and its shape depends strongly on the asymmetry of the three contacts. These conclusions characterize the low energy 
physics and are unaffected by electron-hole decoherence in the junction. At large energies, the density of states 
depends strongly on decoherence, as in two terminal junctions. 
Our most important result is the possibility to efficiently switch between a regime where the spectrum of states in the junction 
shows a separation that can be as large as $|\Delta|$, to a regime where the neighboring states are separated by the smallest energy 
scale of the junction, the level spacing in the normal metal. This opens interesting opportunities for designs of single fermion 
manipulation schemes and for the realization of Majorana bound states in multi-terminal superconducting junctions.
We propose a tunneling transport experiment that can observe the predicted results and calculate its measurement output.

\begin{acknowledgements}

We appreciate the useful discussions with B. Dou\c{c}ot, R.-P. Riwar, M. Houzet, J. S. Meyer, and T. Yokoyama.
We acknowledge support from the French National Research Agency (ANR) through the project ANR-Nano-Quartets (ANR-12-BS1000701). 
This work has been carried out in the framework of the Labex Archim\`ede (ANR-11-LABX-0033) 
and of the A*MIDEX project (ANR-11-IDEX-0001-02), funded by the "Investissements d'Avenir" 
French Government program managed by the French National Research Agency (ANR).

Yu. V. acknowledges the support of  the Nanosciences Foundation in Grenoble, in the framework of its
Chair of Excellence program.

\end{acknowledgements}

\appendix*
\section{Exact numerical method}

For the exact numerical solution we have used the matrix representation of Green's functions in the full Keldysh-Nambu space, rather than spectral vectors as in the main text. The Green's functions of the superconductors are
\begin{align}
\check{G}_i &= \Mab{\check{G}^R_i}{\check{G}^K_i}{0}{\check{G}^A_i} ;\\
\check{G}^R_i &= \frac{1}{\xi}\Mab{\epsilon}{\Delta_i}{-\Delta_i^*}{-\epsilon} ;\\ 
\check{G}^A_i &= -\frac{1}{\xi^*}\Mab{\epsilon^*}{\Delta_i}{-\Delta_i^*}{-\epsilon^*} ;\\
\check{G}^K_i &= (\check{G}^R_i-\check{G}^A_i)\tanh(E/2k_BT) .
\end{align}
where complex energies have been introduced $\epsilon=E+i0^+$ and $\xi=\sqrt{\epsilon+|\Delta|}\sqrt{\epsilon-|\Delta|}$. 

The transport properties are encoded into the current matrices flowing from reservoir $i$,
\begin{align}
I_{i} = \frac{2_se^2}{\pi\hbar}\sum_p \frac{ T_{p,i} [\check{G}_i,\check{G}]}{4+T_{p,i}(\{\check{G}_i,\check{G}\}-2)}.
\end{align}
where $\check{G}$ denotes the Keldysh-Nambu Green's function of the node.

Decoherence is described by a matrix current flowing into a fictitious terminal,
\begin{align}
I_{f} &= \frac{2_se^2}{\pi\hbar}\sum_{i,p} T_{p,i} [\check{G}_f,\check{G}_c] ,\\
\check{G}_f &= -i \frac{E\tau_d}{\hbar} \Mab{\tau^3}{0}{0}{\tau^3},
\end{align}
where $\vec{\tau}$ denotes the vector of Pauli matrices in Nambu space. The choice of $\check{G}_f$ is such that $I_{f} $ carries no particle or heat currents, but only the information about loss of electron-hole coherence.

The unknown Green's function $\check{G}$ is determined by the balance of current matrices, equivalent to the extremization of the action \ref{eq:action} defined in the main text. 
\begin{align}
\sum_i \check{I}_{i} + \check{I}_{f}= 0 .
\end{align}
It is convenient to express the current balance equation as a commutator between matrix $\check{G}$ and a matrix $\check{M}$ that depends on all $\check{G}_i$ and on the unknown matrix $\check{G}$,
\begin{align}
[\check{G},\check{M}] &= 0, \\
\check{M} &= \displaystyle{\sum_{i=\{1,2,3\}}\sum_p}\ T_{p,i}\frac{\check{G}_i}{4+T_{p,i}(\{\check{G}_i,\check{G}\}-2)}+\notag\\ 
\ &\ \displaystyle{\sum_{i=\{1,2,3\}}\sum_p}\ T_{p,i} \check{G}_f .
\end{align}

The above equations can be solved numerically using the following iterative algorithm \cite{Yuli, Mihajlo}.

1. Start with an initial guess for $\check{G}$.

2. Find the matrix $\check{P}$ that brings both $\check{G}$ and $\check{M}(\check{G})$ in diagonal form, i.e. $\check{P}$ contains the eigenvectors of $\check{M}$ as its columns.

3. Diagonalize the matrix $\check{M}$, $\check{M}^{\prime}=\check{P}^{-1}\check{M}\check{P}$.

4. Update the Green's function of the node $\check{G}^{\rm new}=\check{P}\ {\rm sgn}\left[{\rm Re}\left(\check{M}^{\prime}\right)\right]\ \check{P}^{-1}$.

5. Repeat iteratively until convergence, i.e. matrices $\check{G}$ and $\check{G}^{\rm new}$ are within a predefined accuracy.

6. Repeat the iterative procedure for every point in energy space.

The solution at a given point in energy space provides a convenient initial guess for the iterative procedure at the next step in energy.


\begin{thebibliography}{100}

\bibitem{Golubov} A. A. Golubov and M. Yu. Kuprianov, J. Low Temp. Phys. \textbf{70}, 83 (1988).
\bibitem{Kitaev} A. Yu. Kitaev, Phys. Usp. \textbf{44}, 131 (2001).
\bibitem{Fu} L. Fu, C. L. Kane, Phys. Rev. Lett. \textbf{100}, 096407 (2008).
\bibitem{Anton} B. van Heck, S. Mi, and A.R. Akhmerov, Phys. Rev. B \textbf{90}, 155450 (2014).
\bibitem{Riwar} R.-P. Riwar, M. Houzet, J. S. Meyer, and Yu. V. Nazarov, preprint arXiv:1503.06862 (2015).
\bibitem{Tomohiro} T. Yokoyama and Yu. V. Nazarov, preprint arXiv:1508.00146 (2015).
\bibitem{meandYuli} C. Padurariu and Yu. V. Nazarov, Europhys. Lett. \textbf{100}, 57006 (2012).
\bibitem{Carlo} C.W. J. Beenakker and H. van Houten, in \textit{Single-Electron Tunneling and Mesoscopic Devices}, edited by H. Koch and H. L\"ubbig (Springer, Berlin, 1992).
\bibitem{Schon} W. Belzig, C. Bruder, and G. Sch\"on, Phys. Rev. B \textbf{54}, 9443 (1996).
\bibitem{Yuli} Yu. V. Nazarov, Superlattices Microstruct. \textbf{25}, 1221-1231 (1999).
\bibitem{Yulivectors} Yu. V. Nazarov, Phys. Rev. Lett. \textbf{73}, 1420 (1994).
\bibitem{Gueron} S. Gu\'eron, H. Pothier, N. O. Birge, D. Esteve, and M. H. Devoret, Phys. Rev. Lett. \textbf{77}, 3025 (1996).
\bibitem{Freyn} A. Freyn, B. Dou\c{c}ot, Denis Feinberg, and R. M\'elin, Phys. Rev. Lett. \textbf{106}, 257005 (2011).
\bibitem{Jerome} J. Rech, T. Jonckheere, T. Martin, B. Dou\c{c}ot, D. Feinberg, R. M\'elin, Phys. Rev. B \textbf{90}, 075419 (2014).
\bibitem{Denis} D. Feinberg, T. Jonckheere, J. Rech, T. Martin, B. Dou\c{c}ot, R. M\'elin, Eur. Phys. J. B \textbf{88}, 99 (2015).
\bibitem{Becks} D. Beckmann, H. B. Weber, and H. v. Lohneysen, Phys. Rev. Lett. \textbf{93}, 197003 (2004).
\bibitem{Falci} G. Falci, D. Feinberg, and F. W. J. Hekking, Europhys. Lett. \textbf{54}, 255 (2001).
\bibitem{Belzig} J. P. Morten, A. Brataas, W. Belzig, Phys. Rev. B \textbf{74}, 214510 (2006).
\bibitem{Yulibook} Yu. V. Nazarov and Ya. M. Blanter, Quantum transport: Introduction to Nanoscience, (Cambridge University Press, 2009).
\bibitem{smilegap} J. Reutlinger, L. Glazman, Yu. V. Nazarov, and W. Belzig, Phys. Rev. Lett. \textbf{112}, 067001 (2014).
\bibitem{DelftMajorana} V. Mourik, K. Zuo, S. M. Frolov, S. R. Plissard, E. P. A. M. Bakkers, L. P. Kouwenhoven, Science \textbf{336}, 1003-1007 (2012).
\bibitem{Mihajlo} M. Vanevi\'c, PhD Thesis, (February 2008, Basel).
   
\end{thebibliography}
\end{document}